\documentclass[twocolumn,english,aps,prl]{revtex4}
\usepackage[T1]{fontenc}
\usepackage[latin1]{inputenc}
\usepackage{graphicx}
\usepackage{amssymb}

\makeatletter

\usepackage{graphicx}

\usepackage{graphicx}

\usepackage{babel}
\makeatother
\begin{document}

\title{Universality of quantum critical dynamics in a planar OPO}

\author{Peter D. Drummond$^{1}$ and Kaled Dechoum$^{2}$}

\affiliation{(1) ARC Centre of Excellence for Quantum-Atom Optics, University
of Queensland, Brisbane 4072, Queensland, Australia. \\
 (2) Instituto de F\'{\i}sica da Universidade Federal Fluminense,
Boa Viagem, 24210-340, Niter\'{o}i, Rio de Janeiro, Brazil}

\begin{abstract}
We analyze the critical quantum fluctuations in a coherently driven
planar optical parametric oscillator. We show that the presence of
transverse modes combined with quantum fluctuations changes the behavior
of the `quantum image' critical point. This zero-temperature non-equilibrium
quantum system has the same universality class as a finite-temperature
magnetic Lifshitz transition.
\end{abstract}
\maketitle
The non-equilibrium system consisting of a nonlinear crystal inside
a laser driven Fabry-Perot interferometer, that couples sub-harmonic
intra-cavity modes to a harmonic pump~\cite{Peter}, is known as
an optical parametric oscillator (OPO). As well as demonstrating quantum
squeezing~\cite{Wu} and EPR entanglement in numerous quantum information
experiments, this system is widely used in frequency conversion applications.
In single-mode experiments, there is a critical point in the phase
diagram. This is caused by an increase in the pump intensity, which
results in a transition from a disordered (but quantum squeezed) phase
below threshold, to an ordered phase with a coherent output above
threshold. 

When extended to an interferometer with multiple transverse modes,
more complex dynamical effects occur due to diffraction of the down-converted
light, which are governed by the Swift-Hohenberg equation near threshold\cite{Classical}.
The theory can be quantized\cite{Drummond}, and hence includes quantum
fluctuations\cite{Lugiato}. Experimentally observable\cite{Fabre,Pointer}
`quantum images' are evidence for quantum pattern formation with spatio-temporal
correlations in the output quadratures. Thus, the system can show
both spatial critical fluctuations and non-equilibrium spontaneous
pattern formation, which occurs in many fields of physics and other
sciences\cite{Gollub}. 

In this Letter we apply the theory of finite size scaling to solve
for the critical quantum dynamical properties, and obtain the universality
class this phase-transition corresponds to. We provide a full quantum
description of this non-equilibrium system using the positive P-representation,
focusing on the nature of the critical point and critical fluctuations.
This allows us to obtain an analytic solution for the functional distribution
of the large critical fluctuations caused by quantum noise in the
down-conversion process. There is an unexpected universality property
in the solutions. Even though this is a non-equilibrium quantum system
of coupled boson fields, we find the in one quadrature, the quantum
fluctuations have exactly the same behavior as a classical thermal
system of fields at a two-dimensional Lifshitz point, which is a model
commonly used to describe the phase transition to a modulated magnetic
phase; in the complementary quadrature there is strong entanglement.

The unitary evolution of the OPO system can be described by the Hamiltonian\cite{Lugiato}\begin{eqnarray}
\widehat{{\cal H}} & = & \sum_{n=0,1}\int d^{2}\vec{x}\left\{ \hat{A}_{n}^{\dagger}\left[\omega_{n}-\frac{v^{2}}{2\omega_{n}}\nabla^{2}\right]\hat{A}_{n}\right\} \nonumber \\
 &  & +i\hbar\int d^{2}\vec{x}\hat{A}_{0}\left\{ \mathcal{E}^{*}e^{2i\omega_{L}t}-\chi\hat{A}_{1}^{\dagger2}\right\} \,\,.\end{eqnarray}

The term $\chi$ is a coupling parameter that depends on the nonlinear
crystal, the frequencies of the field modes are $\omega_{1}$, $\omega_{0}=2\omega_{1}$,
$v$ is the intracavity group velocity, and $\hat{A}_{n}$ is the
$n-th$ photon field. The pump is described by the amplitude ${\mathcal{E}}$
that could carry a spatial structure - but here we will assume a constant
plane wave input. In addition, there are damping effects due to output
couplings from the cavity mirrors, which can be well approximated
using as a Markovian master equation for the density matrix $\hat{\rho}$,
so that:\begin{equation}
\frac{\partial\hat{\rho}}{\partial t}=\frac{1}{i\hbar}\left[\widehat{{\cal H}},\hat{\rho}\right]+\sum_{n=0,1}\gamma_{n}{\cal L}_{n}\left[\hat{\rho}\right]\,,\end{equation}
where ${\cal L}_{n}\left[\hat{\rho}\right]=\int d^{2}\vec{x}\left[2\hat{A}_{n}\hat{\rho}\hat{A}_{n}^{\dagger}-\hat{\rho}\hat{A}_{n}^{\dagger}\hat{A}_{n}-\hat{A}_{n}^{\dagger}\hat{A}_{n}\hat{\rho}\right]$
describes the output coupling from the $n$th intra-cavity mode, with
damping rate $\gamma_{n}$. This leads to a set of Fokker-Planck equations,
mapped from the the quantum density matrix, using operator representation
theory. These are valid provided boundary terms vanish in the mapping
transformation, which we have checked numerically. Using the positive
P-representation, we derive the following stochastic equations\cite{Peter,Lugiato,Fabre}
in a rotating frame at frequency $\omega_{L}$:

\begin{eqnarray}
\frac{\partial A_{1}}{\partial t} & = & -\widetilde{\gamma}_{1}A_{1}+\chi A_{1}^{+}A_{0}+i\gamma_{1}D\nabla^{2}A_{1}+\sqrt{\chi A_{0}}\xi_{1}(t,\vec{x})\,\,.\nonumber \\
\frac{\partial A_{0}}{\partial t} & = & -\widetilde{\gamma}_{0}A_{0}+{\mathcal{E}}-\frac{\chi^{*}}{2}A_{1}^{2}+\frac{i\gamma_{1}}{2}D\nabla^{2}A_{0}\,\,.\label{Peq}\end{eqnarray}

\noindent Here we write the two dimensional Laplacian causing diffraction,
as $\nabla^{2}=\partial^{2}/\partial x^{2}+\partial^{2}/\partial y^{2}$.
The complex relaxation rates are $\widetilde{\gamma}_{i}=\gamma_{i}(1+i\Delta_{i})$.
The relative detunings between the pump laser at $2\omega_{L}$, and
the modes supported by the cavity are $\Delta_{0}=(\omega_{0}-2\omega_{L})/\gamma_{0}$,
and $\Delta_{1}=(\omega_{1}-\omega_{L})/\gamma_{1}$ . The diffraction
rate is defined as $D=v^{2}/(2\gamma_{1}\omega_{1})$. The stochastic
field $\xi_{1}$ which describes quantum noise is real and Gaussian,
with correlations of $\left\langle \xi_{1}(t)\right\rangle =0$ and
$\left\langle \xi_{1}(\vec{x},t)\xi_{1}(\vec{x}^{\prime},t^{\prime})\right\rangle =\delta^{2}(\vec{x}-\vec{x}^{\prime})\delta(t-t^{\prime})$. 

In addition, there are equations that correspond to the hermitian
conjugate fields. As elsewhere in this Letter, we obtain these by
conjugating the constant terms and replacing the stochastic and noise
fields according to: $A_{i}\rightarrow A_{i}^{+}$, $\xi_{1}\rightarrow\xi_{1}^{+}$,
where $\xi_{1}$, $\xi_{1}^{+}$ are independent real Gaussian noises.
These two noise fields are sufficient to generate all quantum effects,
and are physically caused by the discrete nature of the photon pairs
produced in down-conversion. The c-number fields $A_{i}(t,\vec{x}),\, A_{i}^{+}(t,\vec{x})$
are therefore not complex conjugate, although they are stochastically
equivalent in terms of normally-ordered operator moments to photon
operator fields $\widehat{A}_{i}(t,\vec{x}),\,\widehat{A}_{i}^{\dagger}(t,\vec{x})$.
Thus for example, the photon number density is$\langle\widehat{A}_{i}^{\dagger\,}(t,\vec{x})\widehat{A}_{i}(t,\vec{x}')\rangle=\langle\, A_{i}^{+}(t,\vec{x})A_{i}(t,\vec{x'})\rangle\,\,.$

If we remove the transverse modes from the above equations we return
to the well known single mode OPO theory. This system has a quantum
critical point - a phase transition in the infinite volume limit,
where the quantum fluctuations are reduced below the vacuum level
for the squeezed quadrature and become huge for the unsqueezed quadrature\cite{Plimak}.
In a recent analysis~\cite{CDD} of this problem near the critical
point, going beyond the linear theory, we obtained a scaling law for
the squeezing quadrature spectrum near threshold, and the parameters
for the optimum squeezing. 

The introduction of transverse modes generates a spatial structure
in the sub-harmonic field with an intensity correlation function known
as a {}``quantum image''~\cite{Lugiato}, since it is supported
by quantum fluctuations. To treat this problem analytically, we can
perform an adiabatic elimination of the stable pump mode in the limit
of $\gamma_{0}\gg\gamma_{1}$ and $\Delta_{0}\rightarrow0$. That
is, we assume that the pump mode has a short relaxation time. 

Neglecting pump diffraction - which is negligible in the critical
regime - we obtain an adiabatic solution for the pumped field: $\overline{A}_{0}=\left({\mathcal{E}}-\chi^{*}A_{1}^{2}/2\right)/\gamma_{0}$,
together with a similar equation for the conjugate term. This solution
takes into account the depletion of the pumping mode that supplies
energy for down-converted light, and leads to an adiabatic equation
for the down-converted field: 

\begin{eqnarray}
\frac{\partial A_{1}}{\partial t} & = & -\widetilde{\gamma}_{1}A_{1}+\frac{\chi}{\gamma_{0}}\left({\mathcal{E}}-\frac{\chi^{*}}{2}A_{1}^{2}\right)A_{1}^{+}\nonumber \\
 & + & i\gamma_{1}D\nabla^{2}A_{1}+\sqrt{\chi\overline{A}_{0}}\xi_{1}(t,\vec{x})\,\,.\label{Adiabatic}\end{eqnarray}

Next, we introduce dimensionless variables $\tau=t/t_{0}$ and $\vec{r}=\vec{x}/x_{0}$,
with a corresponding down-converted field $\alpha=x_{0}A_{1}$. Due
to critical slowing down, the characteristic length $x_{0}$ and time
$t_{0}$ scale as $t_{0}=1/(g\gamma_{1})$ and $x_{0}^{2}=D/\sqrt{g_{c}}$,
where the effective nonlinear coefficient is $g_{c}=|\chi|^{4/3}/[8D\gamma_{0}\gamma_{1}]^{2/3}$
, and we will assume that $g_{c}\ll1$. The dimensionless driving
field is $\widetilde{\mu}=\chi{\mathcal{E}}/[\gamma_{1}\gamma_{0}]=\mu+i\theta$.
We also introduce appropriately scaled noise fields with $\xi(\tau,\vec{r})=x_{0}\sqrt{t_{0}}\xi_{1}(t,\vec{x})$,
and the corresponding hermitian conjugate terms. 

With these definitions, we find that:\begin{eqnarray}
\frac{\partial\alpha}{\partial\tau} & = & \frac{1}{g_{c}}\left[-(1+i\Delta_{1})\alpha+\left(\widetilde{\mu}-4g_{c}^{2}\alpha^{2}\right)\alpha^{+}\right]\nonumber \\
 & + & \frac{1}{\sqrt{g_{c}}}\left[i\nabla_{r}^{2}\alpha+\xi\sqrt{\left(\widetilde{\mu}-4g_{c}^{2}\alpha^{2}\right)}\right]\,.\label{scaled}\end{eqnarray}
This equation includes both the down-conversion term proportional
to $\alpha_{1}^{+}$, which generates a squeezed signal --- together
with a nonlinear saturation term proportional to $-\alpha_{1}^{2}\alpha_{1}^{+}$,
which limits the down-converted amplitude, and leads to finite size
critical fluctuations. 

Critical fluctuations are most usefully analyzed with scaled quadratures
that correspond to experimentally accessible homodyne detection. These
are defined as \begin{eqnarray}
X(\tau,\vec{r}) & = & \sqrt{g_{c}}\left[\alpha_{1}(\tau,\vec{r})+\alpha_{1}^{+}(\tau,\vec{r})\right]\nonumber \\
Y(\tau,\vec{r}) & = & i\left[\alpha_{1}^{+}(\tau,\vec{r})-\alpha_{1}(\tau,\vec{r})\right]\;\;.\label{quad}\end{eqnarray}

Similarly, there are quadrature noise fields defined as $\xi_{x}\,(\tau,\vec{r})=\xi(\tau,\vec{r})+\xi^{+}(\tau,\vec{r})$,
and $\xi_{y}\,(\tau,\vec{r})=i\sqrt{{g_{c}}}\left[\xi^{+}(\tau,\vec{r})-\xi(\tau,\vec{r})\right]$.
The resulting quantum dynamical equations for these signal field quadratures
are:\begin{eqnarray}
\frac{\partial X}{\partial\tau} & = & -\left[\gamma_{x}+X^{2}+{g_{c}}Y^{2}\right]X-\left[\gamma_{xy}+\nabla^{2}\right]Y+\xi_{x}\nonumber \\
{g_{c}}\frac{\partial Y}{\partial\tau} & = & -\left[\gamma_{y}+{g_{c}}X^{2}+{g_{c}}^{2}Y^{2}\right]Y-\left[\gamma_{yx}-\nabla^{2}\right]X+\xi_{y}\,\,.\nonumber \\
\label{eq:XYequn}\end{eqnarray}
 The linear decay matrix that couples the $X$ and $Y$ quadratures
is given by:\begin{equation}
\left[\begin{array}{cc}
\gamma_{x} & \gamma_{xy}\\
\gamma_{yx} & \gamma_{y}\end{array}\right]=\left[\begin{array}{cc}
(1-\mu)/{g_{c}} & \,-(\theta+\Delta_{1})/\sqrt{{g_{c}}}\\
(\Delta_{1}-\theta)/\sqrt{{g_{c}}} & (1+\mu)\end{array}\right]\,\,.\label{decayrate}\end{equation}

We assume also that close to threshold, and for small enough detunings,
$\gamma_{x}=O(1)$ and $\gamma_{xy}=-(\theta+\Delta_{1})/\sqrt{g_{c}}=O(1)$.
We can always choose quadrature phases so that $\theta=\Delta_{1}+O(g_{c})$.
With this choice, $Y$ is mainly coupled to the $X$ quadrature via
the diffraction term, which couples noise from the critical fluctuations
back into the squeezed quadrature. This implies that $\tilde{\mu}=1+O(g_{c})$
and $\gamma_{y}=2+O(\sqrt{g_{c}})$, so that the noise correlations
are given by;\begin{equation}
\langle\xi_{x}(\tau,\vec{r})\xi_{x}(\tau',\vec{r}')\rangle=2\delta(\tau-\tau')\delta^{2}(\vec{r}-\vec{r}')+O(g_{c})\,\,.\end{equation}
We can now perform a second type of adiabatic elimination, which is
valid in a neighbourhood of the critical point. This takes into account
the fact that the fluctuations in the $X$ quadrature become very
slow near threshold, while the $Y$ quadrature still responds on fast
time-scales of order $1/\gamma_{1}$. To leading order we can drop
terms of $O(\sqrt{g_{c}})$ where $g_{c}\ll1$, and approximate the
above equations as follows: \begin{eqnarray}
\frac{\partial X}{\partial\tau} & = & -\gamma_{x}X-\gamma_{xy}Y-X^{3}-\nabla^{2}Y+\xi_{x}\nonumber \\
0 & = & -2Y+\nabla^{2}X\,\,.\label{X-elim}\end{eqnarray}
 We can therefore eliminate the fast or non-critical quadrature variable
$Y$, by writing the steady state solution of the $Y$ quadrature
as $Y\simeq\nabla^{2}X/2$ . This produces a reduced equation for
the critical quadrature variable $X$, which is valid near threshold:
\begin{equation}
\frac{\partial X}{\partial\tau}=-\gamma_{x}X-X^{3}-\frac{\gamma_{xy}}{2}\nabla^{2}X-\frac{1}{2}\nabla^{4}X+\xi_{x}\,\,.\label{eq:GL}\end{equation}

The above Langevin equation is a Ginzburg-Landau equation describing
the critical quadrature dynamics. Unlike the usual application of
this equation, we note that system is a non-equilibrium one. The noise
term $\xi_{x}$ is of quantum origin rather than thermal origin, and
is present at zero temperature. It is possible to write an equivalent
functional Fokker Planck equation for the probability density $P[X]$,\begin{equation}
\frac{\partial P}{\partial\tau}=\frac{\delta}{\delta X}\left[\left(\gamma_{x}+X^{2}+\frac{\gamma_{xy}}{2}\nabla^{2}+\frac{1}{2}\nabla^{4}\right)X+\frac{\delta}{\delta X}\right]P\,\,,\label{landauFPE}\end{equation}
 and look for the equilibrium distribution in the form $P[X]=Nexp(-{V}[X])$,
where ${V}(X)$ is a potential functional. Making this substitution,
the solution for the distribution $P[X]$ is given by : \begin{equation}
P\propto e^{\left[-\int d^{2}\vec{r}\left(2\gamma_{x}X^{2}+X^{4}-\gamma_{xy}[\nabla X]^{2}+[\nabla^{2}X]^{2}\right)/4\right]}\,\,.\end{equation}

This expression is exactly the same as the Ginzburg-Landau free energy
of a next nearest neighbor magnetic interaction, where $X$ plays
the role of an order parameter. That is, we have been able to map
this problem into a soluble magnetic phase-transition equation with
a Lifshitz point\cite{Hornreich}. The phase diagram of this optical
system should therefore have two ordered phases, one of them a spatially
modulated phase associated with a pattern formation. This generic
behavior is known to occur in an OPO, from previous analysis\cite{Classical}. 

In this analogy, the {}``optical paramagnetic phase'' corresponds
to a random photon emission from the OPO operating below threshold,
the {}``optical ferromagnetic phase'' to a continuum emission uniformly
distributed in the transverse plane parallel to the cavity operating
above threshold. In the {}``optical ferromagnetic modulated phase'',
we have a continuum emission but with modulated quadrature in this
plane. At the Lifshitz point, all three phases co-exist. 

The line $\gamma_{x}=0$ is the line of the second-order phase transition
between order-disorder (coherent-incoherent) states. In the incoherent
phase below threshold, $\gamma_{x}>0$, and in the uniform coherent
phase above threshold $\gamma_{x}<0$, as expected in the single mode
case. If $\gamma_{x}$ vanishes we have a Lifshitz point over the
line $\gamma_{x}=0$, and thus a triple point characterizing the coexistence
of the three phases. This holds in the case of perfect tuning of the
signal field inside the cavity, so that $\Delta_{1}=0$. 

In condensed matter physics, the nature of the Lifshitz point\cite{Kaplan,Hornreich79}
is crucially dependent on the order parameter and spatial dimension.
Depending on the dimensionality of the order parameter, the system
may or may not have a true phase transition in two dimensions. Our
system has a one-dimensional real order parameter and two spatial
dimensions with transverse modes, so we expect that this system should
have a true phase transition in the infinite volume limit at finite
temperature. According to the Mermin-Wagner theorem\cite{MerminWagner},
increasing the order parameter dimension (as in type II down-conversion)
would result in phase fluctuations that completely destroy any long
range order. Similarly, reducing the spatial dimension would result
in a continuous transition without a threshold. 

\begin{figure}
\includegraphics[%
  width=6cm]{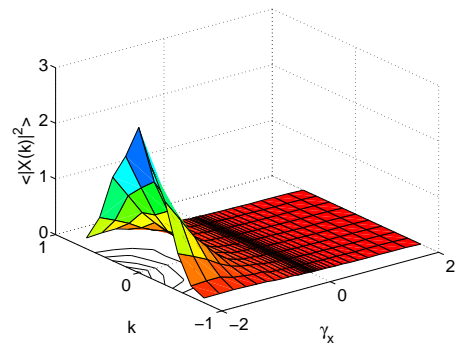}

\caption{Variance of the quadrature Fourier component $\langle|X(k)|^{2}\rangle$
as a function of $\gamma_{x}$. Results obtained on a $100\times100$
lattice with a $40\times40$ domain size, using periodic boundary
conditions and averaging over $100$ stochastic trajectories.}
\end{figure}

A numerical simulation of the Ginzburg-Landau equations (\ref{eq:GL})
with a continuously scanned input shows that the sub-harmonic quadrature
correlations appear to have a true critical point in two transverse
dimensions. This result is shown in Figure (1), which graphs $\langle|X(k)|^{2}\rangle$
around $k=0$, as a function of the driving field $\gamma_{x}$ near
threshold. 

While these large fluctuations are occurring, we note that there are
still strong non-classical correlations in the squeezed quadrature.
This can be seen by analysing the relevant equations to the next order
in $g_{c}$, which we also simplify by using a Gaussian factorization\cite{Stochdiagram}
:

\begin{equation}
{g_{c}}\frac{\partial Y}{\partial\tau}\approx-\gamma_{y}Y+\nabla^{2}X-{g_{c}}Y\langle X^{2}\rangle-\tilde{\gamma}_{yx}X+\xi_{y}\,\,,\end{equation}
where $\tilde{\gamma}_{yx}=\gamma_{yx}+2{g_{c}}\langle XY\rangle$.
In general, one can always choose an optimum local oscillator phase
so that $\theta=k^{2}+\Delta_{1}+2{g_{c}}\langle XY\rangle$, in order
to minimise the feedback of critical fluctuations into the squeezed
quadrature at a given transverse momentum $k$. This leads to Fourier
solutions which showing that entanglement\cite{DrummondFicek} between
the modes of momentum $k$ and $-k$ can still occur at small enough
wave-vectors, resulting in a universal squeezing spectrum as a function
of frequency:\begin{equation}
V(\Omega)=1-\frac{1-{g_{c}}(\langle X^{2}\rangle+\gamma_{x})}{({g_{c}}\Omega/2)^{2}+\left(1+{g_{c}}(\langle X^{2}\rangle-\gamma_{x})/2\right)^{2}}\,\,.\end{equation}

This result differs from the linearised predictions of earlier treatments\cite{Lugiato}.
A graph of the resulting spectrum in the Gaussian approximation is
shown in Fig (2), compared to the linearized squeezing spectrum, showing
large differences near threshold.

\begin{figure}
\includegraphics[%
  width=6cm]{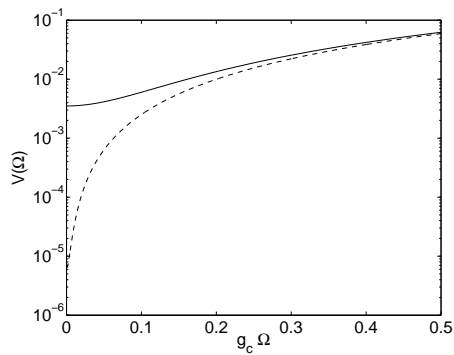}

\caption{Squeezing spectrum as a function of frequency, with (solid line)
and without (dotted line) nonlinear corrections. The parameters used
are $\gamma_{x}=0.5$, $g_{c}=0.01$.}
\end{figure}

In summary, we have shown that the planar non-equilibrium OPO with
quantum noise can be mapped to a magnetic phase-transition in two
dimensions. Since the present case has a scalar, real order parameter
it is analogous to the uni-axial ($m=1$) magnetic order parameter
case, which is known to have a thermal equilibrium Lifshitz-point
phase-transition at finite temperature\cite{Hornreich79}. This demonstrates
a striking resemblance between known thermal equilibrium phase-transitions,
and a quantum non-equilibrium system in which quantum noise replaces
thermal noise. In this system there are also quantum correlations
of the emitted photons, causing quantum squeezing and entanglement.
Nevertheless, this highly non-classical behavior is found only in
the squeezed ($Y)$ quadrature which has no critical slowing down
- and co-exists with a rather classical and universal critical fluctuation
field in the conjugate ($X$) quadrature. 

PDD acknowledges support from the Australian Research Council.

\end{document}